\documentclass[10pt]{article}
\usepackage{graphicx,amssymb, amstext, amsmath, epstopdf, booktabs, verbatim, gensymb, geometry, appendix, lmodern,bchart,pgfplots,url}
\usepackage{chngcntr}
 
\counterwithout{figure}{section}
\counterwithout{table}{section}

\pgfplotsset{width=7cm,compat=1.8}

\newcommand*\Title{KOALA Project Report 1}
\newcommand*\cpiType{--- KOALA Project Report 1}
\newcommand*\Date{February 2018}

\title{What privacy concerns do parents have about children's mobile apps, and how can they stay SHARP?}
\author{Jun Zhao \\Ulrik Lyngs\\Nigel Shadbolt}
\date{\today}

\usepackage{cpi} 

\begin{document}

\begin{titlepage}
\maketitle
\end{titlepage}

\linespread{1.15} 

\begin{executive}

Tablet computers are widely used by young children. According to the 2016 ``Children and Parents' Media Use and Attitudes'' report (Ofcom), children aged 5 to 15 years are, for the first time, \textbf{spending more time online than watching TV}. A 2017 update of the same report shows that parents are becoming more concerned about their children's online risks compared to the previous year. 

Parents are working hard to protect their children's online safety. An increasing number of parents are setting up content filtering at home or having regular discussions with their children regarding online risks. However, although risks related to Social Media platforms or social video sharing sites (like YouTube) are widely known, \textbf{risks posed by mobile applications or games (i.e. `apps') are less known}. 

Behind the cute characters, apps used by children can not only have the possibility of exposing them to age-inappropriate content or excessive in-app promotions, but may also \textbf{make a large amount of their personal information and online behaviour accessible to third party online marketing and advertising industry}. Such practices are not unique to children's apps, but young children are probably less capable of resisting the resulting personalised advertisements and game promotions. Currently there are no effective ways to stop these tracking behaviours on mobile devices. However, there are things that parents/families can do by making more informed choices of apps.

In this report, we present findings from our online survey of 220 parents with children aged 6-10, mainly from the U.K. and other western countries, regarding their privacy concerns and expectations of their children's use of mobile apps. Parents play a key role in children's use of digital technology, especially for children under 10 years old. Recent reports have highlighted parents' lack of sufficient support for choosing appropriate digital content for their children. Our report sheds some initial light on parents' key struggles and points to immediate steps and possible areas of future development.




\frame{
\textbf{Three key findings}
\begin{enumerate}
	\item Parents are generally concerned about their children's online privacy, however, when choosing mobile apps for their children, \textbf{parents primarily focus on the content of the apps and what the apps do, instead of the personal information that might  be collected by the apps}.
	\item Most parents use a range of technical restrictions to safeguard their children, however, \textbf{apps used by the children, as reported by the participant families, are not always  appropriate for their age even though parents seem to be vigilant}. 
	\item Although parents are fairly happy with their children's awareness of privacy risks, and a good proportion of them also regularly discuss privacy issues with their children, \textbf{parents struggle with persuading their children to choose alternative apps when risks arise}.
\end{enumerate}
}

These findings imply that we need to \textbf{continue presenting \textit{specific guidance} to parents} in order to support their choice of digital content for their young children.~Further, we need to look more deeply into whether current guidance on apps' age appropriateness is sufficiently useful for parents, and how parents can be better supported to mediate the choice of digital content by involving their children.~Finally, we need to \textbf{better understand the implications of personal data tracking by mobile apps for the development and well-being of young children}. Evidence in this space could be critical to pushing for \textit{policies and practices} encouraging more transparency about personal data collection by apps designed for children.


\end{executive}

\section{Introduction}

Tablet computers (e.g. iPads or Android tablets) are widely used to complement education and provide entertainment at home and school. In the UK, 44\% of children aged between 5 to 11 have their own tablet computers or use them as the major means to go online~\cite{ofcom2016}. Tablet computers are used in over 70\% of UK schools to complement education. A similar trend is visible in the US, where ownership of tablets by young children has grown five-fold between 2011 and 2013 \cite{commonsense2013}. This is transforming the learning experience for young children~\cite{judge2015using,papadakis2016comparing} and can potentially complement classroom teaching~\cite{neumann2014touch,schacter2016improving}.

However, these technologies raise new privacy risks and challenges to families and parenting~\cite{livingstone2014eu}:
\begin{itemize}
\item Firstly, parents often find it hard to keep up with rapid developing technologies. As they have not themselves grown up with social media or mobile technologies, parents themselves are often the ones oversharing information online or making blind choices of technologies~\cite{minkus2015children}. 
\item Secondly, parents often believe that their children are too young to understand online privacy risks and delay the discussion with them. This leaves young children mostly under their parents' protections, and lacking the essential skills to fend themselves or seek help when needed~\cite{kumar2018cscw}.
\item Finally, technologies are not making parenting in the digital age particularly easy. The large number of applications (`apps') that can be downloaded for free are largely supported by monetisation of user's personal information~\cite{acquisti2016economics,kummer2016private}. While the use of cute characters may suggest benign practices, \textbf{a large amount of personal information and online behaviour may be collected from children's apps and shared with third party online marketing and advertising industry}~\cite{reyes2017our}. Such practices are not unique to children's apps, but currently there are not effective ways to detect or ban these tracking behaviours on mobile devices. A follow-up report in 2015 by the US Federal Trade Commission revealed \textbf{a continued existence of personal tracking by mobile apps for children}\footnote{https://www.ftc.gov/news-events/blogs/business-blog/2015/09/kids-apps-disclosures-revisited}. Our and related studies have shown that apps from the `family app' store, a section of the Google Play store designed for families, have a comparably high number of third-party trackers as game or news apps designed for everyone, with \textbf{over 1 in 4 apps being linked to more than ten different tracking companies}~\cite{reyes2017our,webscience2018}.
\end{itemize}


It is largely agreed that parents these days need better support for safeguarding their children's online safety, and for making informed choices about digital content consumed by their children~\cite{digital2018}. This report provides a first-step understanding about what parents currently struggle with, so that we can \textbf{design and develop the exact kind of support parents most need, and encourage a positive co-learning process for both parents and their children}. As a result of recognising several key knowledge gaps, \textbf{we propose a set of concrete steps that parents can take today to keep a '\textcolor{red}{S}\textcolor{orange}{H}\textcolor{yellow}{A}\textcolor{green}{R}\textcolor{blue}{P}' eye on their children's choice and usage of mobile apps}.

The report is based on an online study focused on parents of children aged 6 to 10. Being the first generation growing up with modern mobile technologies, children in this age group are rapidly developing independent digital literacy skills, while their awareness of online safety is largely under studied. Unlike children in their early teenage years, these children are still maintaining a close relationship with their parents, which may make active interventions or guidances better received than older children~\cite{digital2017}. We recognise that our findings may therefore be limited to this age group. We conclude the report by a brief discussion of the implications of the study and its relation to challenges faced by parents of children in other age groups.
\pagebreak

\section{Research Method}

Our report is based on an online survey conducted in late summer 2017. Participants were recruited through the Prolific Academic system (http://prolific.ac), a crowdsourced online survey platform that is capable of recruiting participants fitting specific criteria. In our survey, participants were required to be parents of, or guardians to, at least one child aged between 6 and 10, who has regular access to a tablet computer or smartphone. Participants came from varied levels of family income and residential areas.  

Questions in the survey were based on findings from our previous qualitative interviews with parents and children from 12 families, and related literature. The goal of the survey was to reach out to a wider demographics, and understand the following:

\begin{itemize}
\item How are mobile apps installed on children's or families' mobile devices? What kind of technical restriction mechanisms or processes are set up by parents to protect children?
\item What kind of privacy concerns do parents have when choosing apps  for their children? How do they impact on parents' actual choice of apps for children?
\item How well do parents know about existing privacy safeguarding tools on mobile devices, such as kids app stores or privacy permission settings? Do they find them effective?

\end{itemize}



The online survey received 250 responses. 29 responses were excluded because the children were outside the age range of 6 to 10. Of the remaining 221 responses, the average age of the parents was 35.9, and 69.5\% were female. Most of our respondents (78\%) were UK residents, 15\% US residents, and the remaining 7\% resided in other western countries. The UK residents were evenly distributed in different regions of the country. The respondents had on average 2.27 children, and the average age of the child selected for completing the survey was 7.91. 

Details about the methodology, including the survey design and additional data tables can be made available upon request.

\pagebreak
\section{Key findings}

\subsection{Children's access to apps are largely under parents' control}
\begin{figure}[h!]
\centering
  \includegraphics[width=0.8\columnwidth]{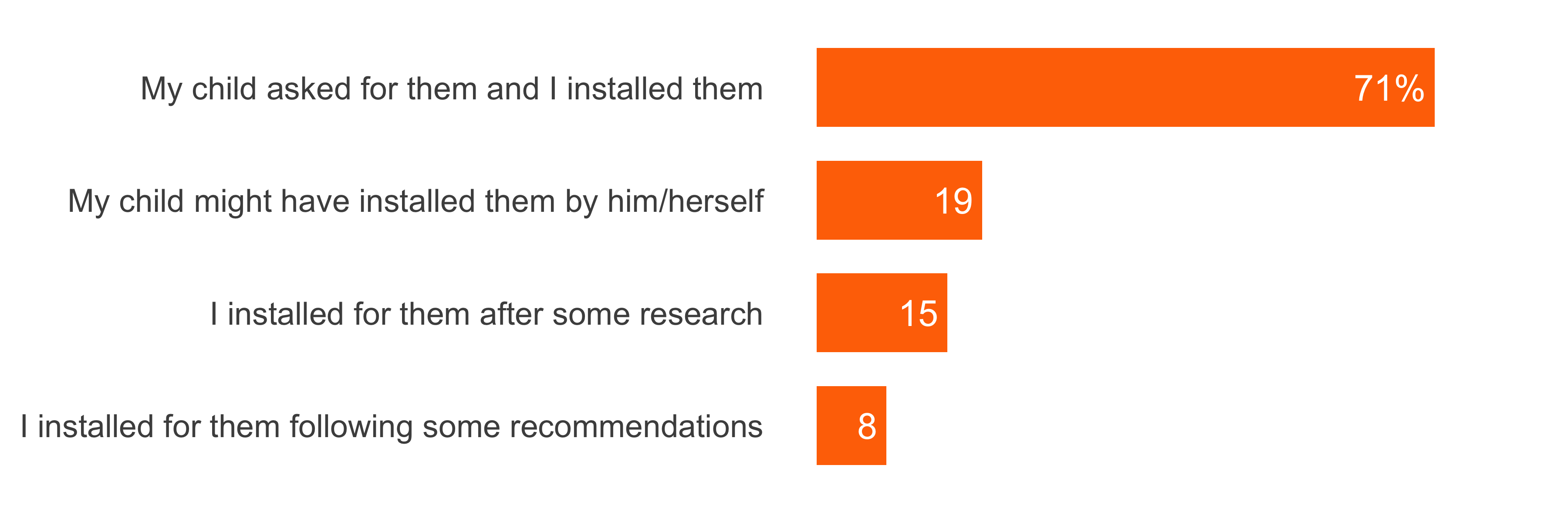}
  \caption{How did your child get these apps installed on the device? (Q15)}~\label{fig:install}
\end{figure}

\begin{figure}[h!]
\centering
  \includegraphics[width=0.8\columnwidth]{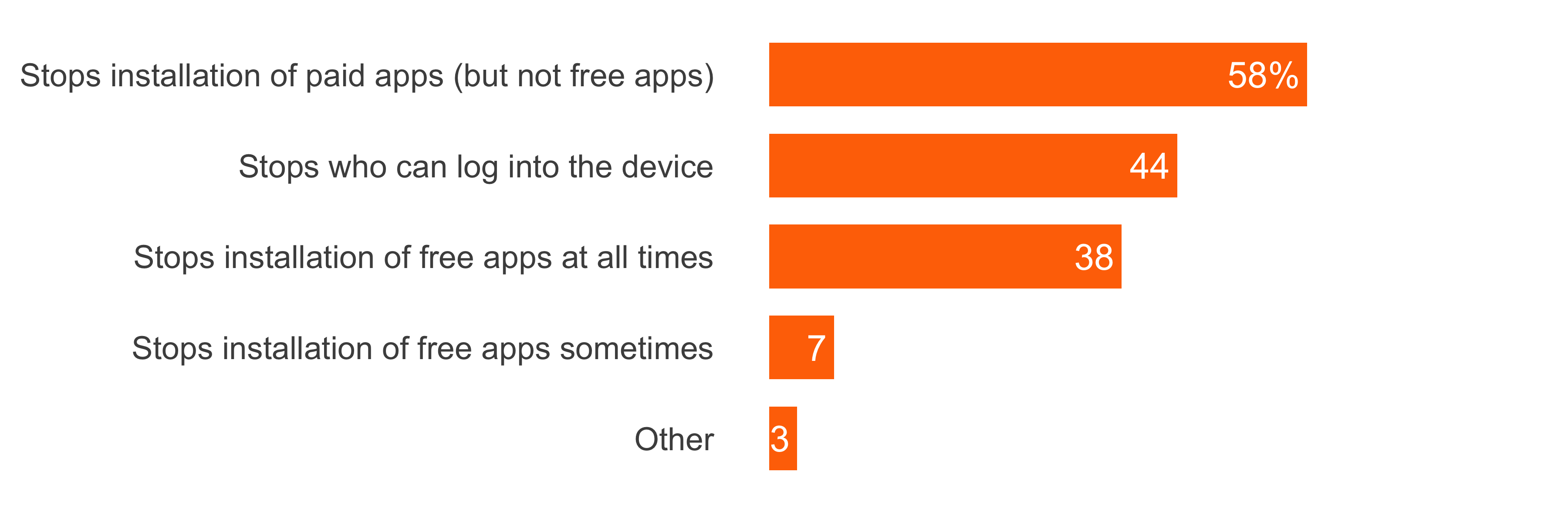}
  \caption{What does the password protection do? (Q17-18)}~\label{fig:password}
\end{figure}

In the survey, parents were asked to report two of their child's favorite mobile apps and how they were installed (Figure~\ref{fig:install}).
\begin{itemize}
\item 71\% of parent respondents said that they had installed the apps after the child had asked for it.
\item 23\% installed the apps after doing some initial research or following recommendations.
\item However, \textbf{a sizable minority (19\%) reported that the child might have installed them by themselves}.
\item Passwords were commonly set up to stop young children from logging into devices by themselves, installing paid apps or any apps. However, \textbf{nearly 24\% did not have any password restrictions on the devices}. 
\item Of those families with password protection, 58\% of them said that the purpose was to stop installation of paid apps, but not of free apps, suggesting that \textbf{their motivation was probably financial rather than privacy-related} (Figure~\ref{fig:password}). 
\end{itemize}

        

\pagebreak

\subsection{Parents expressed a high level of privacy concerns, but content was still king}

The great majority of parents expressed concerns about online privacy in relation to their child's interaction with table devices, with 45\% saying that they `very often' think about it, and 47\% that they `sometimes' or `occasionally' do so. However, \textbf{content} was still the dominant factor for parents when choosing apps to install or deciding to uninstall an app for their young children.

\begin{figure}[h!]
\centering
  \includegraphics[width=0.8\columnwidth]{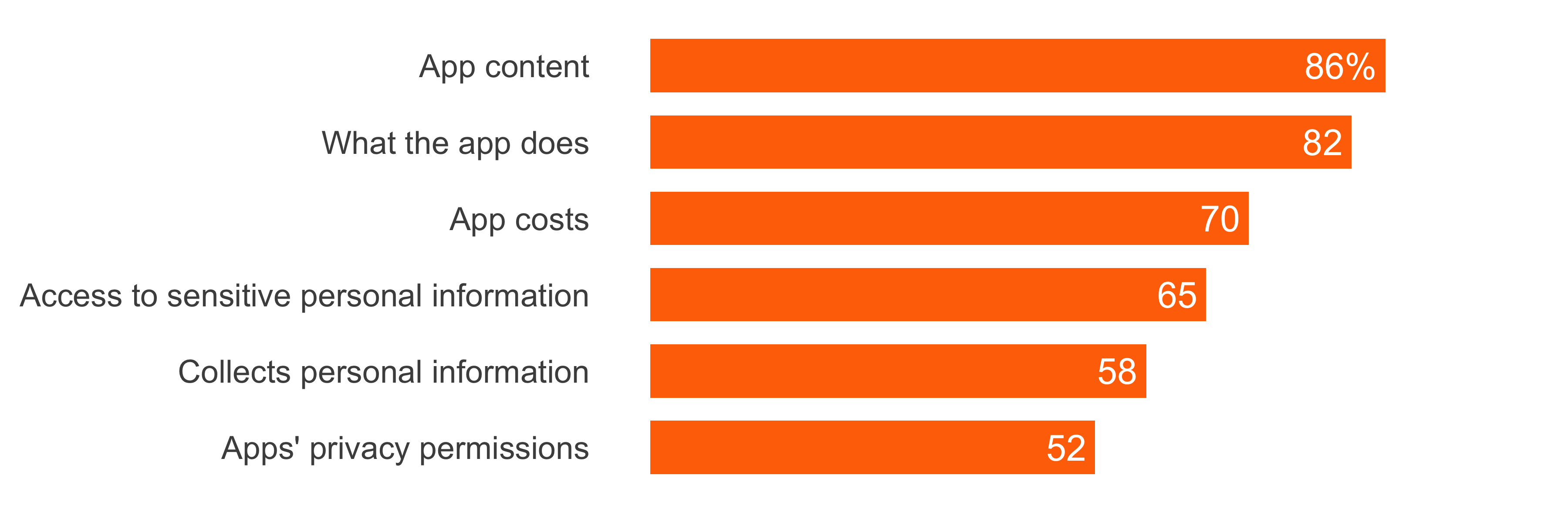}
  \caption{Things considered by the parents when installing apps for their children (Q16)}~\label{fig:consider_install}
\end{figure}

\begin{figure}[h!]
\centering
  \includegraphics[width=0.8\columnwidth]{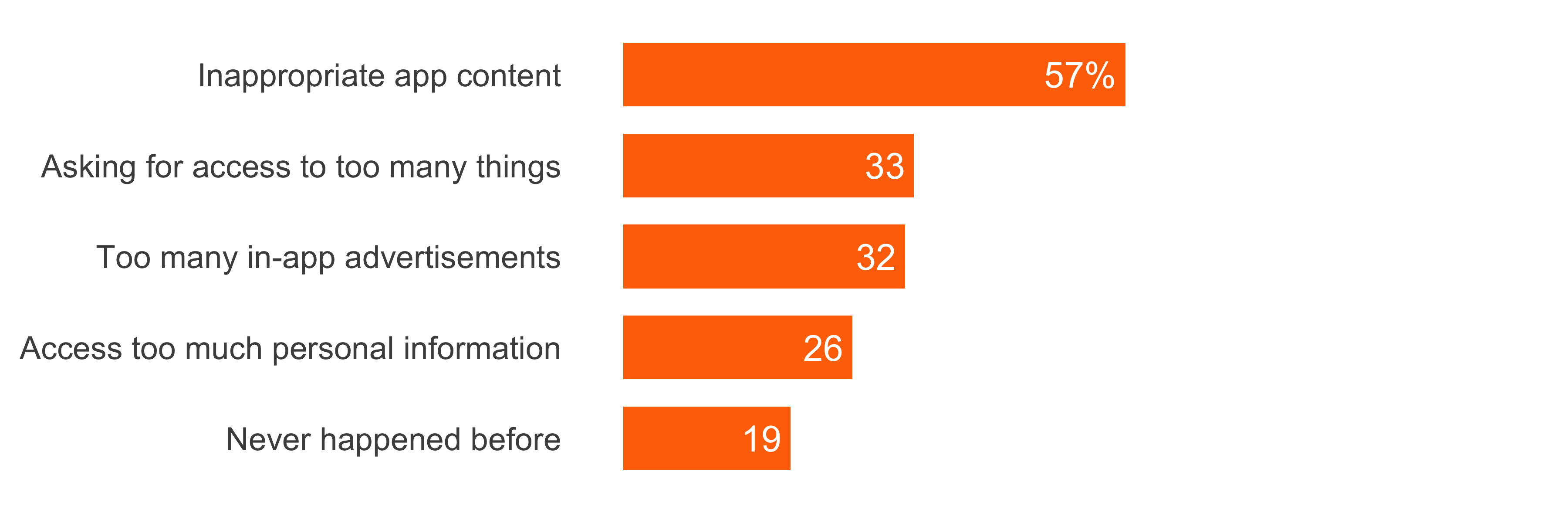}
  \caption{Things considered by the parents when refusing to install or uninstalling an app (Q23)}~\label{fig:consider_uninstall}
\end{figure}

\begin{itemize}

\item When installing new apps for their children, \textbf{more than 4 in 5} parents considered whether the app \textbf{content} was appropriate for their child, while a smaller majority reported considering whether the app collects sensitive information about the child or themselves (65\%), or requires permissions to access particular things on the device, such as camera or location (52\%). Unsurprisingly, most parents also considered the price of the app (70\%, Figure~\ref{fig:consider_install}).

\item More than 3 in 5 parents had refused to install an app (67\%) and more than half of the parents have uninstalled an app (54\%) for their child. The primary reason mentioned by 57\% of parents was that the \textbf{app content was inappropriate}. A smaller number of parents (33\%) said that they had done so because the app asked for access to things not necessary for the app to function, such as camera or location, or specifically that the app was accessing too much information about their child (26\%, Figure~\ref{fig:consider_uninstall}).

\end{itemize}

\pagebreak

\subsection{Children's favorite apps are not always appropriate for their age}

YouTube was the app most frequently mentioned (55\% of all respondents) by the parents as one of their child's two favorite apps, followed by Minecraft (16\%) and Roblox (14\%). 

Table~\ref{tab:apps} shows those of children's favorite apps that were mentioned >5 times in the survey responses:
\begin{itemize}
\item \textbf{5 out 14 apps had an age rating inappropriate for young children aged 6-10} and would require parental guidance\footnote{The information about each app is retrieved from Google Play Store, accessed in December 2017. The UK-specific interpretation of the content rating can be found https://support.google.com/googleplay/answer/6209544?hl=en-GB. PEGI 3-7 are probably more appropriate for young children aged 6-10, than those  rated ``requiring parental guidance''.}, including YouTube as well as some popular social media apps.

\item \textbf{Only 4 apps required no access to sensitive personal data}\footnote{Sensitive personal information refers to the type of information that might reveal a person's identity, such as their unique device ID, location or contact details.}, such as contact details on the device, location information or unique identification of the device.
\end{itemize}

The level of privacy concerns expressed by the parents did not significantly correlate with the type of favorite apps used by the children: parents 'very concerned' about privacy were equally likely to report apps with inappropriate age rating or excessive access to personal information as being among their children's two favourite apps. This implies that \textcolor{orange}{\textbf{parents' privacy concerns do not seem to translate into consistent choice of digital content for their children}}. 

\begin{table}[h!]
    \centering
    \begin{tabular}{c|c|c|c}
     App & Number of mentions & Age rating & Sensitive personal data access \\ \hline
YouTube & 110 & \textbf{Parental guidance} & \textbf{YES} \\
Minecraft & 40 & PEGI 7 & \textbf{YES} \\
Roblox & 29 & \textbf{Parental guidance} & \textbf{YES} \\
Netflix & 14 & \textbf{Parental guidance} & \textbf{YES} \\
CBeebies & 10 & PEGI 3 & NO \\
WhatsApp & 10 & PEGI 3 & \textbf{YES} \\
YouTube Kids & 9 & PEGI 3 & \textbf{YES} \\
Clash Royale & 9 & PEGI 7 & NO \\
Candy Crush & 8 & PEGI 3 & NO \\
Facebook & 8 & \textbf{Parental guidance} & \textbf{YES} \\
Angry Birds & 6 & PEGI 3 & \textbf{YES} \\
Pokemon Go & 5 & PEGI 3 & \textbf{YES} \\
Temple Run & 5 & PEGI 3 & NO \\
Music.ly & 5 & \textbf{Parental guidance} & \textbf{YES}\\
    \end{tabular}
    \caption{Top favorite apps mentioned in the survey (Q14)}
    \label{tab:apps}
\end{table}

\pagebreak

\subsection{Parents' level of concern increased when told about potential implications}
\begin{itemize}
\item While the majority of the parents said that they were concerned about apps accessing their children's camera or location information, \textbf{parents' level of concern increased} when they were presented with the possibility that these permissions might enable strangers to communicate with their child or share their photos (93\%) or that companies might be able to infer sensitive information about their child (90\%). 

\item 24\% parents became more concerned about apps having access to their children's devices' cameras or microphone (see Figure 5).

\item 6\% parents became more concerned about location access (see Figure 6).

\item This implies possibly a lack of understanding about the consequences of having apps access to some critical information about a child's device. \textcolor{orange}{\textbf{Strengthening parents' understanding about the implications of giving apps access to critical information about a device could potentially lead to more informed decisions}}.

\end{itemize}

        

\begin{figure}[h!]
\centering
  \includegraphics[width=0.8\columnwidth]{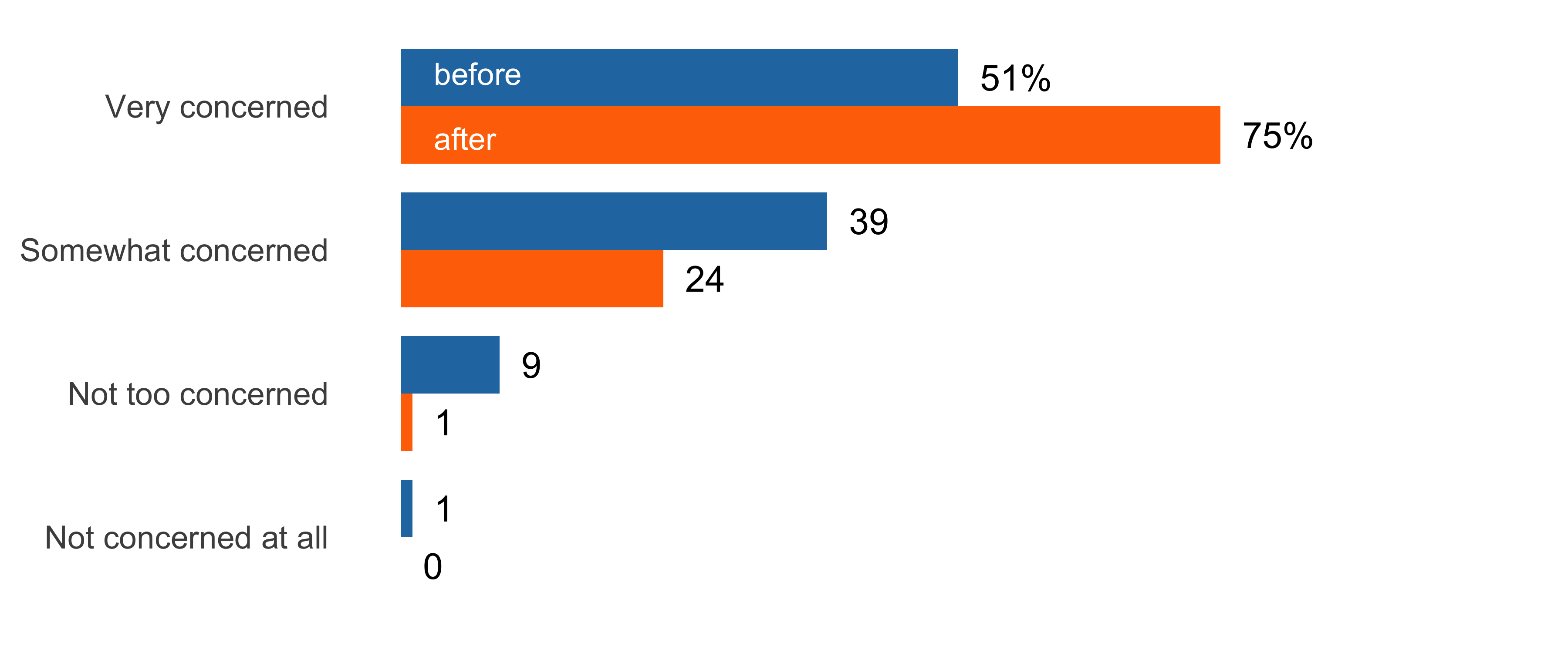}
  \caption{Parents' level of concern regarding access to devices' camera (Q28 vs Q30)}~\label{fig:concern_camera}
\end{figure}


\begin{figure}[h!]
\centering
  \includegraphics[width=0.8\columnwidth]{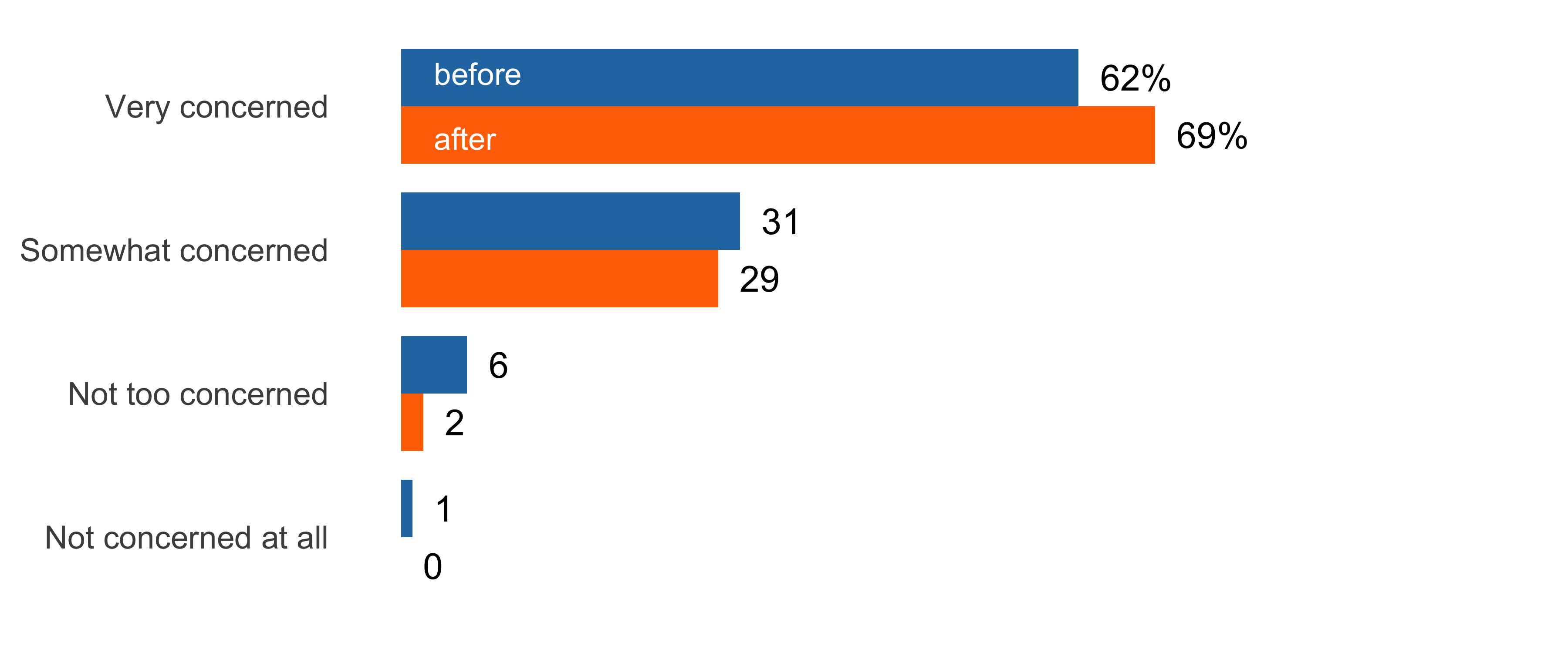}
  \caption{Parents' level of concern regarding access to devices' location information (Q29 vs Q31)}~\label{fig:concern_location}
\end{figure}

\pagebreak

\subsection{Parents know little about kids/family app store provided by leading app markets}

Most parents (61\%) had not heard about Apple's and Google's dedicated sections on their app stores for children and families, in which content ratings are used to group apps into different age categories. Among those who were aware of these stores, the large majority (79\%) used these either `all the time' or `sometimes' (Figure 7) and 66.5\% of them `somewhat' or `strongly' agreed that this reduced their privacy concerns (Figure 8). 

Apple launched its kids store section in September 2013 and Google followed suit in early 2015\footnote{https://www.theguardian.com/technology/2015/may/29/google-android-family-friendly-parents}. These are not separate app stores, but separate sections in the stores to promote apps supposedly more suitable for families and young children, with more controls over in-app promotions and content curation. However, only a small portion of parents know about this resource and find it useful, and our analysis has also shown that these apps are probably not as upfront and transparent about their personal data collection practices as they should~\cite{webscience2018}. \textcolor{orange}{\textbf{This indicates a need to further investigate the actual effectiveness of `kids app stores' for safeguarding young children's online safety and privacy}}.


\begin{figure}[h!]
\centering
  \includegraphics[width=0.8\columnwidth]{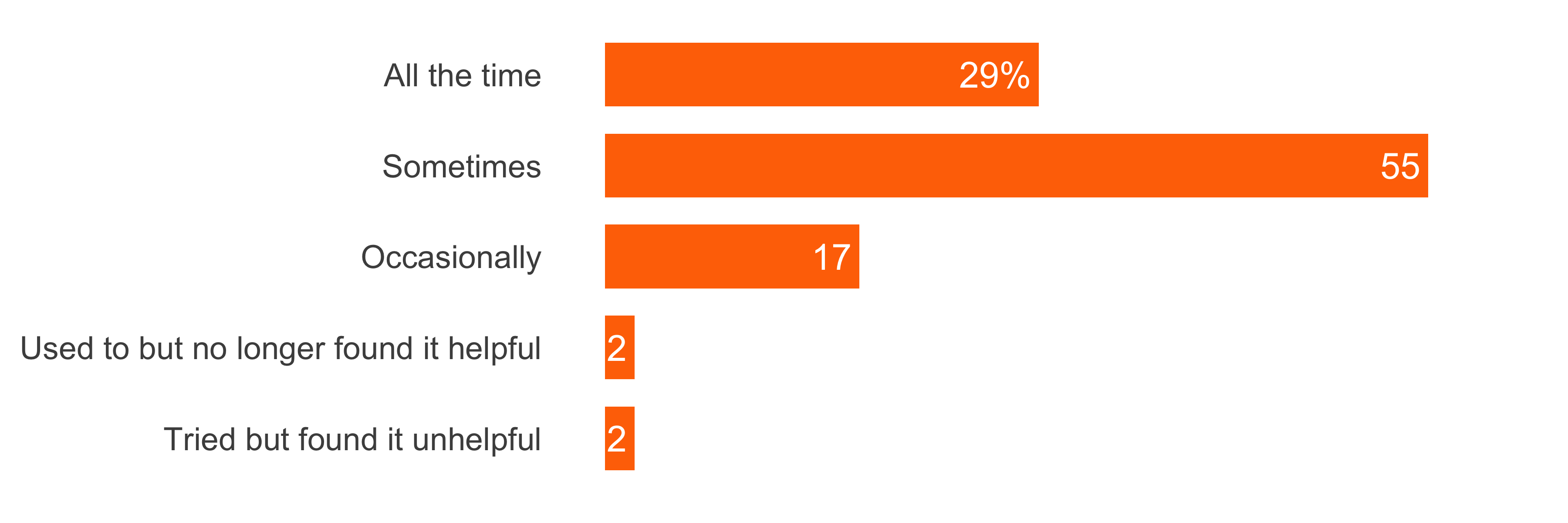}
  \caption{How often kids app stores are used by parents who know about them (Q34-35)}~\label{fig:family_store_use}
\end{figure}

      

\begin{figure}[h!]
\centering
  \includegraphics[width=0.8\columnwidth]{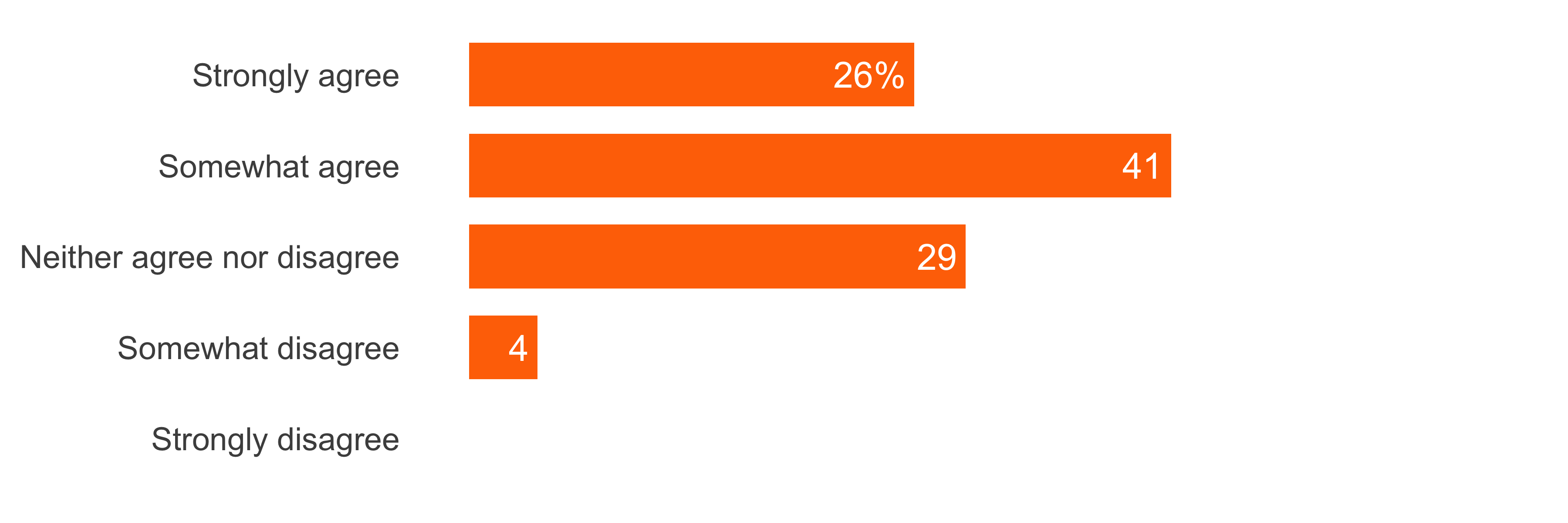}
  \caption{Parents' level of agreement that using kids app stores reduces their privacy concerns (Q36).}~\label{fig:family_store_use_reduce_concern}
\end{figure}

\pagebreak

\subsection{Parents show a good knowledge of privacy permission controls}

\begin{figure}[h!]
\centering
  \includegraphics[width=0.8\columnwidth]{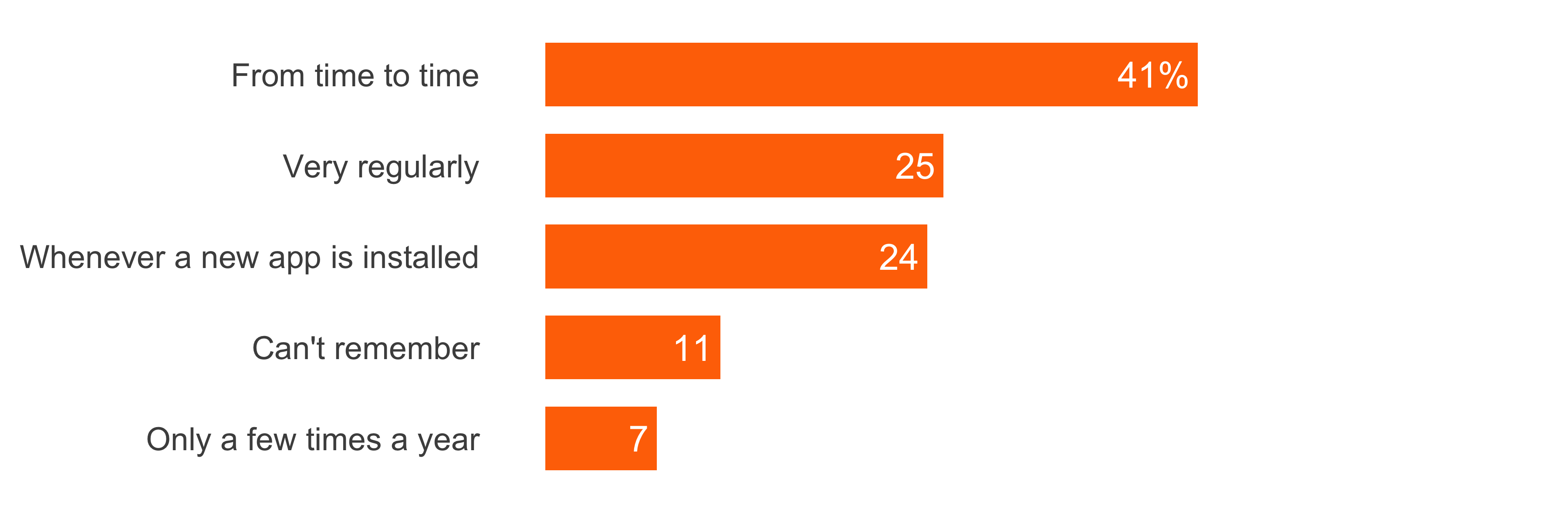}
  \caption{When privacy permission settings were last checked (Q37-38)}~\label{fig:check_permissions}
\end{figure}

\begin{figure}[h!]
\centering
  \includegraphics[width=0.8\columnwidth]{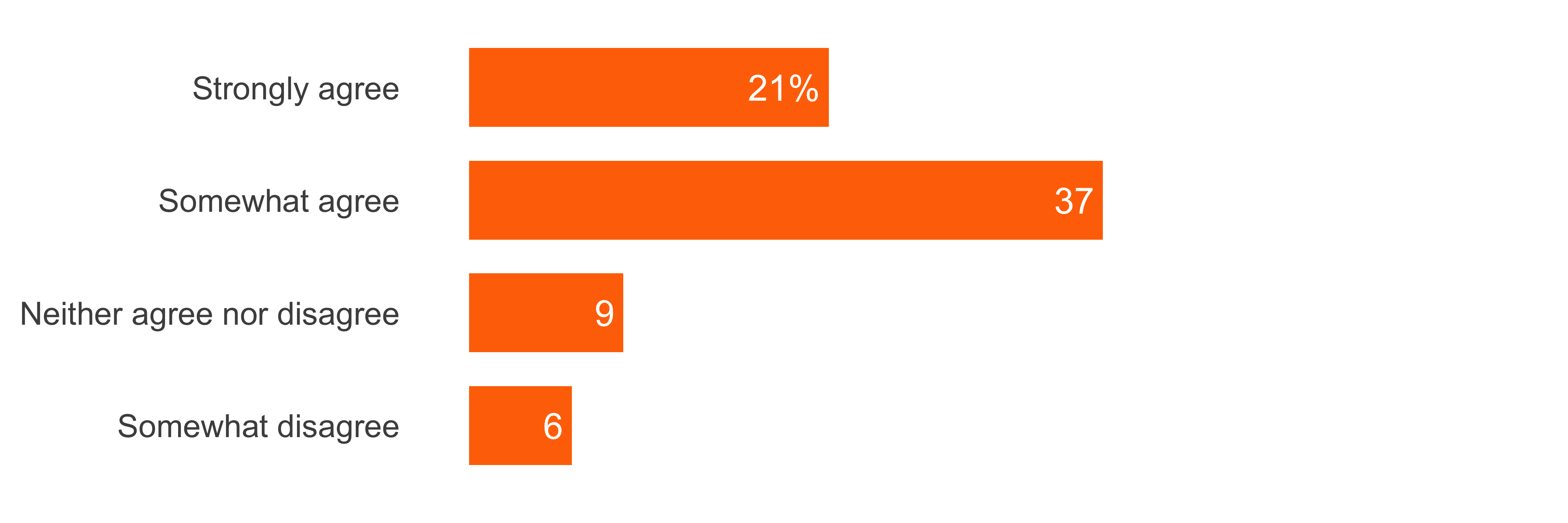}
  \caption{Parents' level of agreement that tuning privacy permission settings reduces their privacy concerns (Q39)}~\label{fig:privacy_perms_reduce_concern}
\end{figure}

\begin{itemize}
\item In contrast to parents' limited knowledge of kids app stores, 72\% of them had heard about app privacy permission settings on mobile devices.

\item Nearly 1 in 4 parents reported that they checked these settings either whenever a new app was installed, and 3 in 5 parents reported that they checked from time to time or very regularly (Figure~\ref{fig:check_permissions}).

\item A minority (16\%) could not remember when they had last checked these settings for their child's device, or only did so a few times a year. 

\item A good majority (78\%) of the parents felt that the settings reduced their privacy concerns about what their child used on the device (Figure~\ref{fig:privacy_perms_reduce_concern}.
\end{itemize}

This positive feedback about privacy controls are encouraging, and contrasts with widespread worry that parents struggle to keep up with mobile technologies. However, our finding 3.4 suggests that parents do not necessarily fully comprehend the implications of different privacy configurations. This is not unique to parents, but a common hurdle for general mobile device users. \textcolor{orange}{\textbf{This implies a further need to raise parents' awareness about the meaning of their choices, and make these implications more meaningful for the family context}}. 


\pagebreak
\subsection{Parents struggle with managing children's emotions when removing inappropriate apps}

\begin{figure}[h!]
\centering
  \includegraphics[width=0.8\columnwidth]{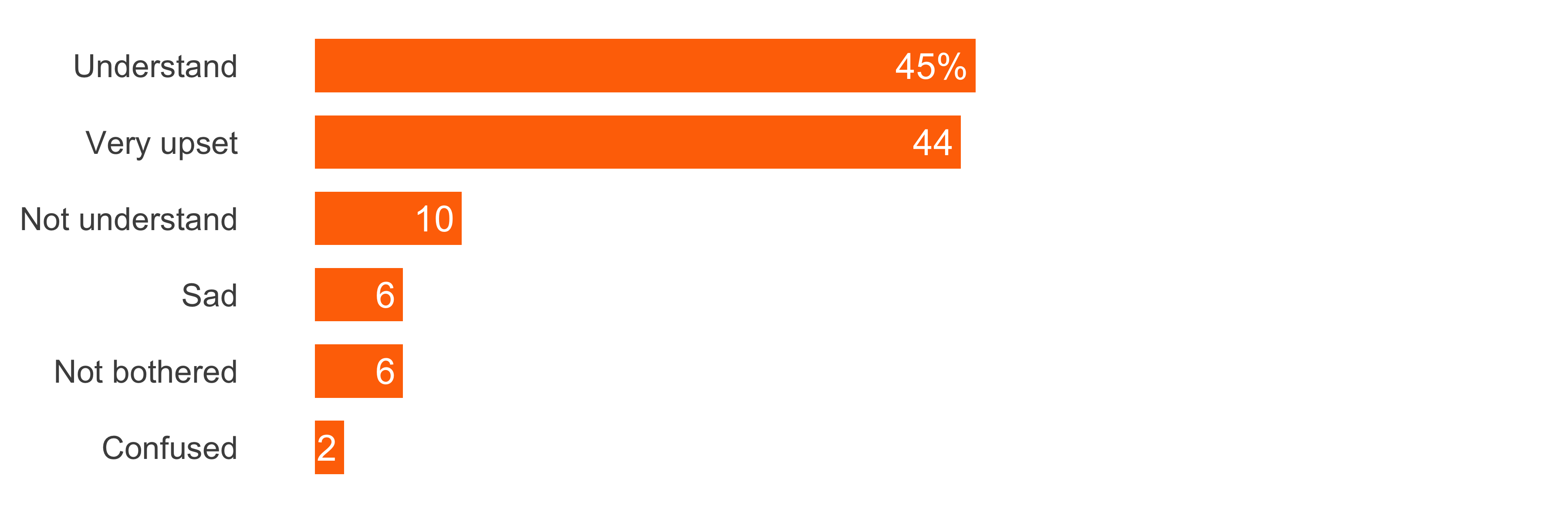}
  \caption{How often kids app stores are used by parents who know about them (Q34-35)}~\label{fig:child_reactions_uninstall}
\end{figure}

Of the 76\% households who reported that they had removed apps from their children's mobile devices due to concerns of privacy risks, many parents said that their children (48\%) were very upset and could not understand their parents' decisions. A minority of parents (7\%) said their children felt sad or confused, and only 6\% said that their children were not bothered. However, still 40\% of the families reported that their children were capable of understanding parents' decisions (Figure~\ref{fig:child_reactions_uninstall}).

Our analysis shows that \textbf{families struggled most to manage this stressful situation when no explanations were given by the parents}. Several parents mentioned that their children could not understand the decision and found it unfair because their friends could still use the apps. \textcolor{orange}{\textbf{This indicates that managing technology usage for parents of young children can be stressful, and the social context lived in by even young children can make digital parenting a challenging task}}.

\vspace{0.5in}


\definecolor{findOptimalPartition}{HTML}{2B83BA}
\definecolor{storeClusterComponent}{HTML}{FF7A33}
\definecolor{dbscan}{HTML}{ABDDA4}
\definecolor{constructCluster}{HTML}{33FFF0}
\definecolor{constructCluster2}{HTML}{808000}
\definecolor{constructCluster3}{HTML}{D7191C}

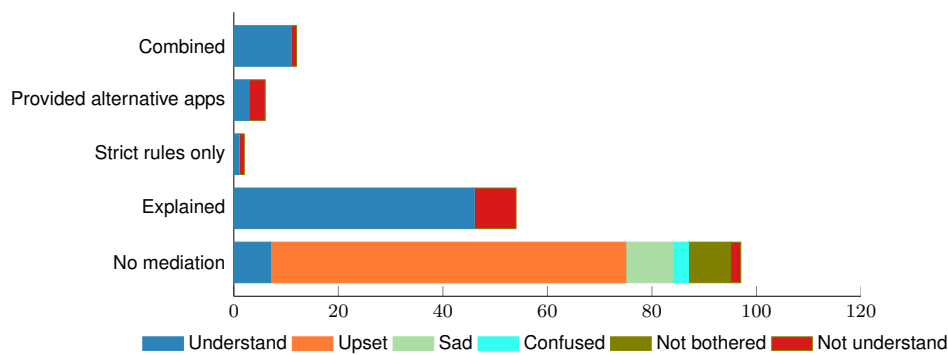
\begin{figure}[h!]
\centering
\scalebox{0.9}{
\begin{tikzpicture}
\begin{axis}[
    xbar stacked,
    legend style={
    legend columns=6,
        at={(xticklabel cs:0.5)},
        anchor=north,
        draw=none
    },
    ytick=data,
    axis y line*=none,
    axis x line*=bottom,
    tick label style={font=\footnotesize},
    legend style={font=\footnotesize},
    label style={font=\footnotesize},
    xtick={0,20,40,60,80,100,120},
    width=.7\textwidth,
    bar width=6mm,
    xlabel={Time in ms},
    yticklabels={No mediation, Explained, Strict rules only, Provided alternative apps, Combined},
    xmin=0,
    xmax=120,
    area legend,
    y=8mm,
    enlarge y limits={abs=0.625},
]
\addplot[findOptimalPartition,fill=findOptimalPartition] coordinates {(7,0) (46,1) (1,2) (3,3) (11,4)};

\addplot[storeClusterComponent,fill=storeClusterComponent] coordinates {(68,0) (0,1) (0,2) (0,3) (0,4)};

\addplot[dbscan,fill=dbscan] coordinates {(9,0) (0,1) (0,2) (0,3) (0,4)};

\addplot[constructCluster,fill=constructCluster] coordinates {(3,0) (0,1) (0,2) (0,3) (0,4)};

\addplot[constructCluster2,fill=constructCluster2] coordinates {(8,0) (0,1) (0,2) (0,3) (0,4)};

\addplot[constructCluster2,fill=constructCluster3] coordinates {(2,0) (8,1) (1,2) (3,3) (1,4)};

\legend{Understand, Upset, Sad, Confused, Not bothered, Not understand}
\end{axis}  
\end{tikzpicture}}
\caption{How young children felt when different mediation approaches were taken by the parents.}
\label{fig:stats}
\end{figure}

\pagebreak

\section{Recommendation to Parents: Keep a `\textcolor{red}{S}\textcolor{orange}{H}\textcolor{yellow}{A}\textcolor{green}{R}\textcolor{blue}{P}' eye}

As a result of our findings, we propose the following concrete steps for parents to consider when managing the choice and usage of mobile apps for their young children:

\begin{itemize}
\item {\fontsize{20}{24}\textcolor{red}{\textbf{S}}}elect: Select apps carefully, by consulting resources like Common Sense Media\footnote{See \url{https://www.commonsensemedia.org/}}.

\item {\fontsize{20}{24}\textcolor{orange}{\textbf{H}}}elp: Talk to your children about asking you for Help when they need it.

\item {\fontsize{20}{24}\textcolor{yellow}{\textbf{A}}}void: Avoid providing any sensitive personal information to the app.

\item {\fontsize{20}{24}\textcolor{green}{\textbf{R}}}ating: Check `Age Rating' of the apps.

\item {\fontsize{20}{24}\textcolor{blue}{\textbf{P}}}rivacy Permission: Check `Privacy Permissions' of the apps.

\end{itemize}

\section{Conclusion}

Parents of children aged 6-10 play a key role in mediating and educating their technology usage. Our survey shows that most parents have good technical restrictions in place on their children's devices, and have a good control of which apps are installed on their children's devices. However, 

\begin{itemize}
\item Parents' concerns about content appropriateness have not always led to consistent choice of apps for children. This indicates a need for further understanding of \textit{why} this is happening and how we may provide better support for the parents.

\item Parents' technical competence can be strengthened with further knowledge about the implications of their technical choices.

\item Parents need better support to manage their struggle of mediating their children's choice of apps~\cite{zaman2016parental}. This is important for not only fostering a positive parenting experience but also for transmitting the essential knowledge and skills to young children, who are at the frontier of risks. 

\item We do not yet fully understand the impact of third party data tracking by mobile apps upon the development and well-being of young children. Evidence collected in this space will be critical to inform policies, regulations and practices~\cite{livingstone2017using}.
\end{itemize}

A recent report on Digital Childhood~\cite{digital2017} provides recommendations for parents, educators and policy makers according to children of different age groups or development stages. In comparison to parents of younger children, parents of children aged 6 to 10 have a great opportunity to \textbf{transition from providing passive protections and controls to proactive mediations}. In comparison to parents of older children, who have more sense of independence, parents of 6-10 year olds have a unique opportunity to develop and maintain a trust relationship with their children, and discuss privacy risks and approaches of addressing them before they reach adolescence. Parents of primary school-aged children are facing rapidly developing new challenges as their children go through  their key development milestones, and parents can play a critical role in fostering their children's future privacy perceptions and behaviours. Their needs demand better understanding and support. 



\section{Acknowledgement}
This research is supported by KOALA (http://SOCIAM.org/project/koala): Kids Online Anonymity \& Lifelong Autonomy, funded by EPSRC Impact Acceleration Account Award, under the grant number of EP/R511742/1.

\section{Contact}
\begin{itemize}
\item Jun Zhao: jun.zhao@cs.ox.ac.uk 
\item Ulrik Lyngs: ulrik.lyngs@cs.ox.ac.uk
\item Nigel Shadbolt: nigel.shadbolt@cs.ox.ac.uk
\end{itemize}

\bibliography{kids-studies,report}
\bibliographystyle{acm}

\end{document}